\documentclass[12pt]{iopart}

\usepackage{graphicx}
\usepackage{dcolumn}
\usepackage{bm}
\usepackage{color}

\begin{document}

\newcommand{\CeCoIn}{CeCoIn$_5$}

\newcommand{\Tc}{$T_\textrm{\footnotesize c}$}

\newcommand{\Hc}{$H_\textrm{\footnotesize c2}$}
\newcommand{\Hczero}{$H_\textrm{\footnotesize c2}(0)$}
\newcommand{\Hcslope}{$H'_\textrm{\footnotesize c2}(\textrm{\Tc})$}
\newcommand{\vF}{$v_\textrm{\footnotesize F}$}

\newcommand{\pc}{$p_\textrm{\footnotesize c}$}
\newcommand{\pmax}{$p_\textrm{\footnotesize max}$}

\definecolor{addcolor}{rgb}{1,0.3,0}
\definecolor{Supcolor}{rgb}{0,0.7,1}
\newlength{\textlarg}
\newcommand{\barre}[1]{%
   \settowidth{\textlarg}{#1}
   #1\hspace{-\textlarg}\rule[0.5ex]{\textlarg}{0.5pt}}

\title{Upper critical field of \CeCoIn}

\author{Ludovic~Howald\footnote{Now at: Physik-Institut der Universit\"at Z\"urich; Winterthurerstrasse 190; CH-8057 Z\"urich; Switzerland},Georg~Knebel, Dai~Aoki, G\'erard~Lapertot, Jean-Pascal~Brison}
\address{
SPSMS, UMR-E CEA/UJF-Grenoble 1, INAC, Grenoble, F-38054, France
}%

\ead{ludovic.howald@physik.uzh.ch, jean-pascal.brison@cea.fr}

\date{\today}

\begin{abstract}
We present a detailed analysis of the upper critical field for \CeCoIn{} under high pressure. We show that, consistently with other measurements, this system shows a decoupling between maximum of the superconducting transition temperature \Tc{} and maximum pairing strength. 
We propose a model in which, to account for the discrepancy in pressure between the maximum of upper critical field and the maximum of \Tc, we introduce magnetic pair breaking effects, already widely suggested by other measurements. We found that within the Eliashberg framwork, the unusual shape of \Hc $(T)$ can be completely reproduced when magnetic pair breaking is taken into account. Surprisingly we obtained that the maximum of pair breaking and of pair coupling coincide in pressure, suggesting that both mechanisms originate from quantum criticality. Our model implies that CeCoIn$_5$ is the first compound of its family that show a clear decoupling between maximum of \Tc{} and quantum criticality.
\end{abstract}

\maketitle

\section{Introduction}
Phase diagrams revealing the range of existence of superconductivity as a function of doping, pressure, or magnetic field and its proximity to an electronic instability (magnetic, charge/spin density wave, metal/insulator transition) are central in the search for the superconducting pairing mechanisms.
This is true for the main families of strongly correlated superconductors like the high-\Tc{} cuprates, iron based pnictides, organics or heavy fermions. But the latter materials are probably the first and best documented cases that evidence a strong relationship between the appearance of magnetic quantum criticality and superconductivity \cite{Mathur1998, Pfleiderer2009}. Of particular importance among heavy fermion superconductors is the so-called 115 family (Ce{\it M}In$_5$), which is often presented as bridging the gap between the (low \Tc) heavy fermions and the high-\Tc{} cuprates, owing to its 2D character, d-wave superconducting states, and pronounced ``non Fermi-liquid'' features \cite{Sidorov2002, Nakajima2007, Knebel2011}.

Among these, \CeCoIn{} can be considered as a paradigm of cerium based heavy fermion superconductors, thanks to its well established proximity to a magnetic instability \cite{Sidorov2002, Kawasaki2003, Sarrao2007, Kenzelmann2008, Sakai2010} and highest \Tc{} of $2.26$ K at zero pressure \cite{petrovic2001}, together with a d-wave \cite{Movshovich2001, Kohori2001, Izawa2001} and multigaps \cite{Seyfarth2008} superconducting state. However, the relation between the field induced quantum critical point (QCP) observed very close to \Hczero{} \cite{Paglione2003, Bianchi2003}, and the pressure  \pmax($\approx1.3$ GPa) of optimal \Tc{}($\approx2.6$ K) remains unsettled \cite{Knebel2010}. \CeCoIn{} is strongly Pauli limited along the $\vec{c}$-axis, as demonstrated by the general behavior of \Hc$(T)$ \cite{Miclea2006}, or the unconventional temperature dependence of the vortex form factor \cite{Bianchi2008}. 
The pressure dependence of the paramagnetic limitation directly give the value of the superconducting gap as $\Delta=g\mu_b$\Hczero{}, and  a maximum of \Hczero{} should correspond to a maximum of \Tc (in BCS $\Delta=1.76k_b$\Tc) in contradiction to the experimental observation \cite{Miclea2006} and to all other comparable known heavy fermions.
So the whole ($T,p,H$) phase diagram of \CeCoIn{} remains puzzling.

In this work, we analyze the pressure dependence of the upper critical field of \CeCoIn{} determined by thermodynamic specific heat measurements, for both in plane and out of plane directions, from measurements ranging from zero pressure up to more than twice \pmax. 
Our analyze show, consistently with other measurements, that \CeCoIn{} is the first example in which there is a clear decoupling between the position under pressure of the maximum of \Tc{} and the maximum of both pairing strength and pair breaking. It suggests a decoupling between the QCP, associated with the latter effects, and the  maximum of \Tc{} revising previous proposals based on transport measurements \cite{Ronning2005}, nevertheless in agreement with predictions for magnetically mediated superconductivity of low dimensional systems \cite{Monthoux1999}.
We propose a new phase diagram for this compound, which puts him further forward as an important paradigm of strongly coupled, 2D, antiferromagnetically mediated superconductor.

\begin{figure}[b]
\begin{center}
\includegraphics[width=1\textwidth]{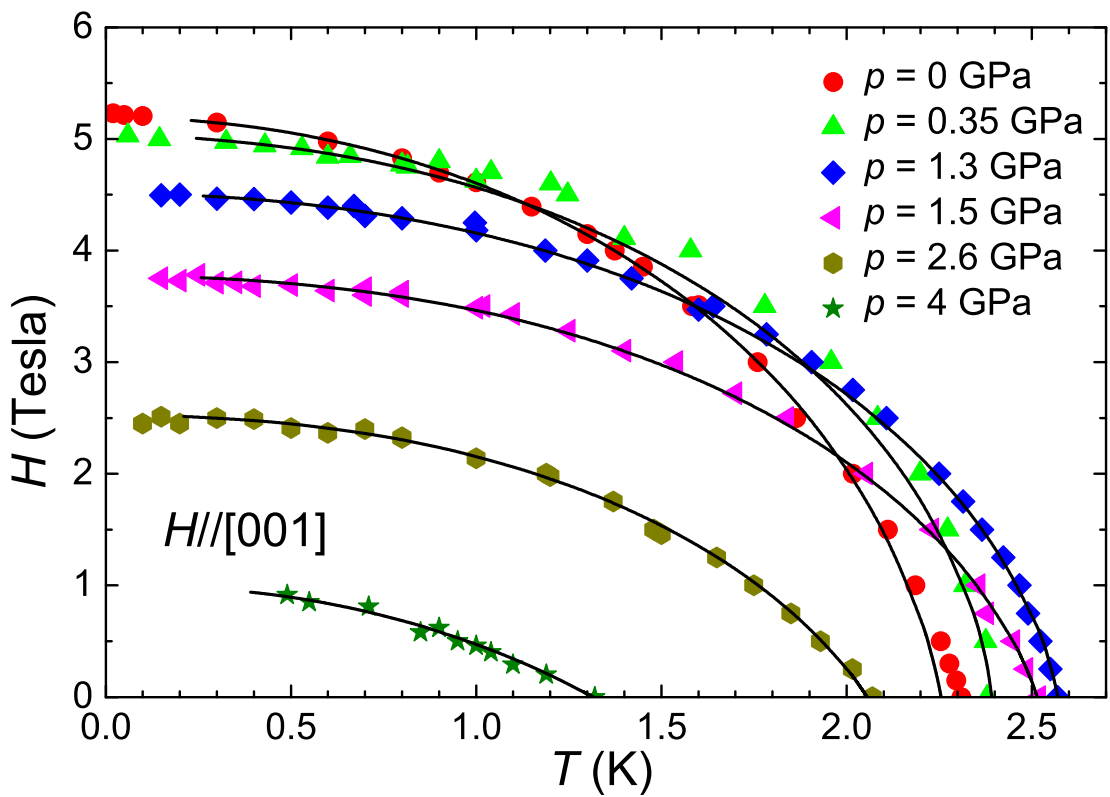}\\

\caption{(colors online) \label{GraHc} Data (points) and fits (full lines) of the upper critical field of CeCoIn$_5$, for $H\parallel [001]$. Experimental data are reported from ac calorimetry as described in \protect{\cite{Knebel2010}}.}
\end{center}
\end{figure}

\begin{figure}[hbt]
\begin{center}\includegraphics[width=1\textwidth]{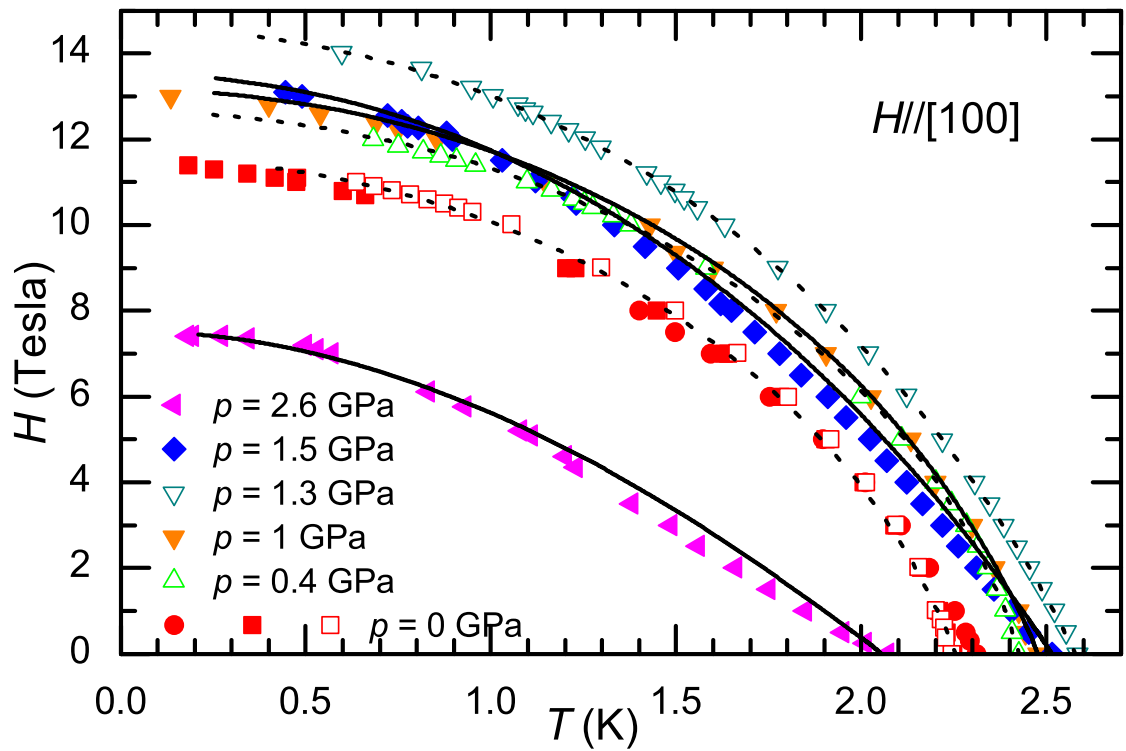}
\caption{\label{GraHa} (colors online) Data from this experiment \protect{\cite{Knebel2010}} (full symbols) and from \protect{\cite{Miclea2006}} (open symbols) of the upper critical field of CeCoIn$_5$ for $H\parallel [100]$. The slope at $T_c$ depends on the measurement technique: at ambient pressure, circles are from specific heat measurements, squares from resistivity. We used the bulk thermodynamic measurements for the fits. Fits (full or dashed lines) as described in the text.}
\end{center}
\end{figure}

\section{Data and analysis}
\begin{figure}[hbt]
\begin{center}
\includegraphics[width=\textwidth]{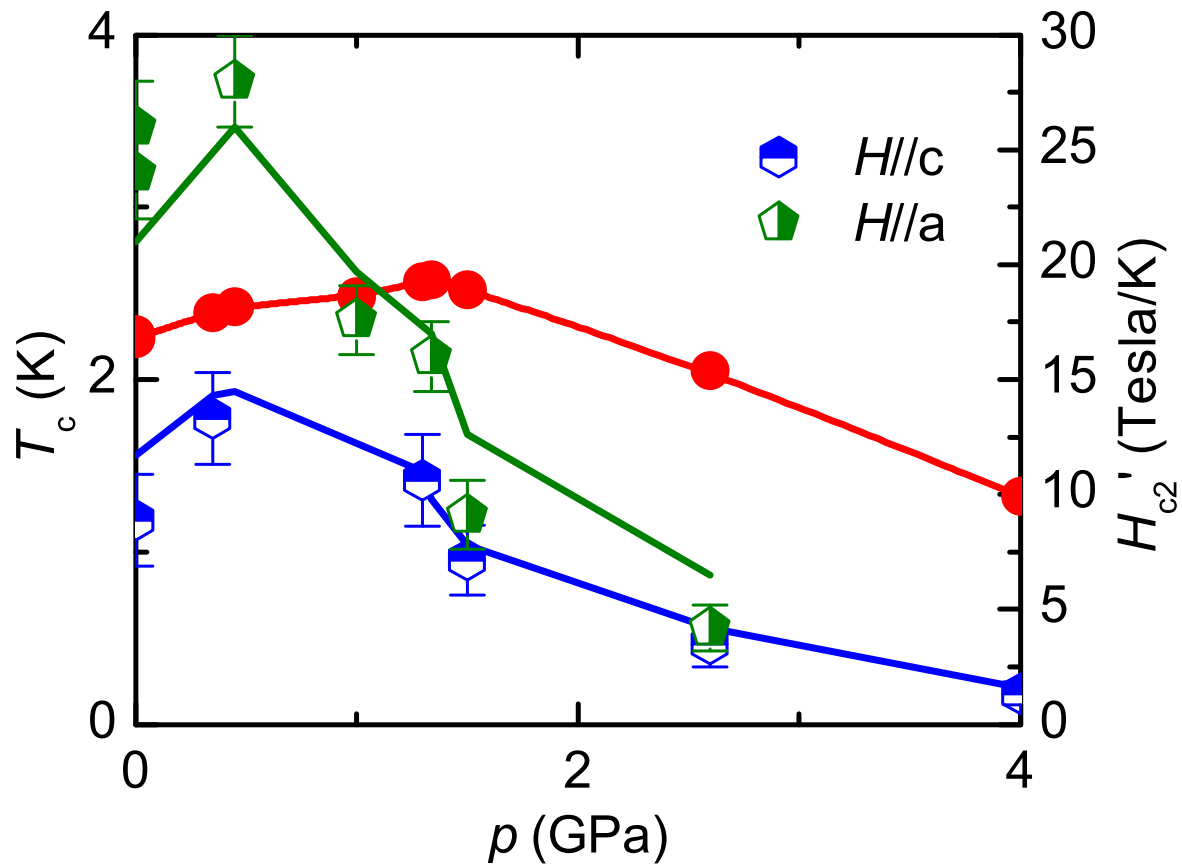}\\
\caption{(colors online) \label{Fig:slope} Pressure evolution of \Tc{} and \Hcslope{} in \CeCoIn. The maximum of \Tc{} is found at $\sim$1.3 GPa while the one of the initial slope of \Hc{} is located at $\sim$ 0.4 GPa. In a conventional picture the two maximums should coincide to reflect a maximum of the coupling interaction. Lines are the values obtained through the model developed in this article (see text).}
\end{center}
\end{figure}

The upper critical field \Hc{} of \CeCoIn, is presented on figures \ref{GraHc} and \ref{GraHa}: Full symbols data were obtained from ac calorimetry measurements ref. \cite{Knebel2010} except data at $p=0, H\parallel \vec{c}$ as well as a curve at $p=0, H\parallel \vec{a}$ (squares) obtained from resistivity measurements, and displayed for comparison. The specific heat data were taken on the same sample in the same diamond-anvil cell, turned $90$ degrees in the fridge for $H\parallel \vec{a}$. Open symbols are from \cite{Miclea2006}. A remarkable feature which can be seen on the raw data of figures \ref{Fig:slope}, \ref{GraHc} and \ref{GraHa} is that, except for the lowest pressure of 0.35 GPa, the initial slope of \Hc{} at \Tc{} (\Hcslope $=(d\textrm{\Hc}/dT)_{T=T_c}$) is continuously lowered with pressure increase, for both field directions (H$\parallel \vec{a}$ and H$\parallel \vec{c}$). \Hcslope{} is controlled only by the orbital limitation, and is proportional to \Tc{} and to the inverse of the square of the Fermi velocity (\vF), for a superconductor in the clean limit. On such a small pressure scale (a few GPa), the evolution of \Tc{} should be governed by that of the coupling strength, usually quantified by a (strong coupling) parameter labeled $\lambda$. The interactions responsible for the pairing do also contribute to the renormalization of the Fermi velocity, precisely by a factor $1/(1+\lambda)$ (see for example: \cite{Crabtree1987}). So if the maximum of \Tc{} in \CeCoIn{} is due to a maximum of $\lambda$, one expects an increase of  $1/$\vF, \Tc{} and \Hcslope{} between $p=0$ and \pmax{}. This is clearly in contradiction with the experimental results. 

A similar problem occurs for the opposite limit of \Hc{}, namely \Hczero{}, for $H\parallel c$. 
The saturating behavior of \Hc{} in \CeCoIn{} at low temperature, for $H\parallel c$ is due to a dominating paramagnetic limitation: $H^P\approx \Delta/g\mu_B$ \cite{Bianchi2002}, where $\Delta$ is the superconducting gap, $g$ the gyromagnetic factor and $\mu_B$ the Bohr magneton.  In the weak coupling limit: $\Delta/\textrm{\Tc}\cong 1.76$ for an isotropic gap, and it is  known that strong coupling effects increase this ratio \cite{Bulaevskii1988}. So an increase of \Tc{} due to an increase of $\lambda$ should enhance $H^P$ beyond the proportionality to \Tc{} \cite{Bulaevskii1988}. Again, this is in strong contradiction with the experimental data of figure \ref{GraHc}, as \Hczero{} is decreasing up to \pmax{}. 

Such a contradiction is unusual among heavy fermion superconductors: most of the time, the pressure variation of \Tc, \Hcslope, and \Hczero{} are fully consistent with the simple expectations given above (see for example \cite{Settai2008}), and can even be quantitatively fitted with essentially only $\lambda$ as a pressure dependent parameter \cite{Glemot1999, Knebel2008}. In particular, for the parent compound CeRhIn$_5$, the situation is very well documented with a maximum of \Tc{} at $p_c \approx 2.4$ GPa corresponding with a maximum of the effective mass as detected by de Haas-van Alphen quantum oscillations, by \Hcslope, or by the $A$ coefficient of the resistivity. A fit of the complete dependence of \Hc$(T)$ with pressure does point to a maximum of $\lambda$ at the maximum \Tc, coinciding with the maximum of the specific heat jump ($\Delta C/C$) at \Tc{} \cite{Knebel2010Butsuri} which is a good measure of the strong coupling effects \cite{Knebel2008}. Moreover, in CeRhIn$_5$, it has been shown that when superconductivity is suppressed by a magnetic field, the antiferromagnetic order is restored with a N\'eel temperature which also vanishes at $p_c\approx 2.4$ GPa \cite{ParkNature2006, Knebel2006PRB}. Therefore, in this system, the pressure ($p_c$) at which the magnetic phase transition occurs at $T=0$ (possible QCP) corresponds to the pressure (\pmax{}) at which the coupling strength and \Tc{} are maximum.

Clearly, for CeCoIn$_5$, such a scenario cannot be applied directly. However, if we put aside the pressure dependence of \Tc, all other results: \Hcslope, \Hczero ($\vec{H}\parallel\vec{c}$), but also the coupling strength as measured by the specific heat jump $\Delta C/C$ \cite{Knebel2010,Sparn2002} or the gap to \Tc{} ratio obtained from nuclear quadrupole resonance \cite{Yashima2004}, are consistent with essentially a decrease of the coupling strength with pressure. They are also consistent with the proposal that under pressure \CeCoIn{} moves away from a QCP. From this standpoint, a natural hypothesis would be that: (i) the pairing strength (measured by $\lambda$) decreases with pressure, (ii) \Tc{} is controlled by $\lambda$ and by an additional (limiting) mechanism. This limiting mechanism should decrease faster with pressure than $\lambda$, so that \Tc{} could reach a maximum a finite pressure, despite the pressure decrease of $\lambda$. The maximum \Tc{} of \CeCoIn{} would then arise ``artificially'', from the competition between the pressure dependence of both mechanisms, and not from a particular critical pressure. We should note that Monthoux and Lonzarich \cite{Monthoux1999} made a full calculation of \Tc{} for d-wave anti-ferromagnetically mediated superconductivity. They obtained that for strong coupling the maximum of \Tc{} is not located at the QCP where the coupling strength is maximum, due to pair breaking associated to magnetic spin fluctuations which grows faster on approaching the QCP \cite{Monthoux1999}. 

Let us elaborate quickly on this point: already for conventional electron-phonon pairing mechanism, it is known that in the strong coupling regime, the presence of thermally excited phonons at \Tc{} is detrimental for superconductivity. For example, the reinforcement of the superconducting properties at low temperatures (like the gap-to-\Tc{} ratio) compared to the weak-coupling case arises from the disappearance  of thermal phonons (as opposed to the ``virtual phonons'' of the pairing mechanism) \cite{Bulaevskii1988}. However, despite the pair-breaking effects of thermal phonons, an increase of the coupling strength will always result in an increase of \Tc{}: it is only the rate of this increase which becomes lower than that of the weak-coupling regime. In the case of magnetically mediated pairing, thermal magnetic fluctuations will have similar pair-breaking effects than thermal phonons. But in addition, for the model mechanism studied by Monthoux and Lonzarich \cite{Monthoux1999}, it is found that for sufficient pair breaking strength, pair-breaking wins over \Tc{} increase. In such a case, approaching the quantum critical point (where the pairing, but also pair breaking is the strongest) may result in a decrease of \Tc.

Evidence for such thermal spin fluctuations in \CeCoIn{} are numerous. For example, presence of a sizeable fraction of fluctuating moments down to the lowest temperatures have been detected in the normal state of \CeCoIn{} \cite{Kohori2001,Kawasaki2003,Yashima2004,Nakatsuji2004}. Coupling between superconductivity and antiferromagnetic (AFM) fluctuations is demonstrated by inelastic neutron scattering, which detects a resonant signal below \Tc{} \cite{Stock2008}. The very large specific heat jump ($\Delta C/C\approx 5$) at the superconducting transition, which is beyond any expectation even for a strong coupling superconductor, could also be explained by the coupling of ``fluctuating paramagnetic moments'' to the superconducting order parameter \cite{Kos2003}. The same model \cite{Kos2003} can also explain the magnetization jump in the mixed state of \CeCoIn{} at \Hc{}. A similar but more general spin-fermion model \cite{Bang2004} also reproduces the specific heat result for magnetic fluctuations in the proximity of a QCP. Kos et al. \cite{Kos2003} could quantify the effect of these magnetic fluctuations (at p=0) from the superconducting transition that the system would have in absence of these fluctuations: $T^{\star}\cong 3T_c$. NQR  \cite{Yashima2004} and residual resistivity \cite{Nicklas2001} under pressure demonstrate also a strong decrease of these AFM fluctuations under pressure, a necessary ingredient for our scenario. Eventually, an extensive study of pair breaking effects by magnetic and non magnetic impurities in \CeCoIn{} \cite{PaglioneNatureP2007} also pointed out the existence of  ``incoherent'' scattering in this compound, supporting the ``two fluids picture'' of \cite{Nakatsuji2004} and the existence of fluctuating moments at very low temperatures. As mentioned in the same paper \cite{PaglioneNatureP2007}, the existence of unpaired electrons in the superconducting state of \CeCoIn{} \cite{Tanatar2005}, or more likely of a very small gap in this compound \cite{Seyfarth2008}, demonstrates a large density of states of low energy excitations, which probably indicates the presence of a sizable pair-breaking mechanism.
 

\section{Model}
In order to extract quantitative information from the pressure dependence of \Hc{}, we used an Eliashberg, s-wave, strong coupling model for its calculation \cite{Bulaevskii1988}. In the spirit of reference \cite{Kos2003}, we added the effect of magnetic impurities to mimic the \Tc{} reduction induced by strong AFM fluctuations. 

Within our model, \Tc{} and \Hc{} are functions of the parameters:
$T_c/\Omega (p)= \Psi(\lambda,\mu^{\star},T_M) $,
$H_{c2} (p,T) = \Phi(T, T_c, \lambda,\mu^{\star}, T_M ,v_F,g)$,
where $\Omega$ is a characteristic temperature of the coupling mechanism (analog to the Debye temperature in the electron-phonon case), $\lambda$ is the strong coupling constant, $\mu^{\star}$ is the coulomb pseudo-potential, fixed to a typical value of 0.1, and $T_M$ gives the characteristic energy of the pair breaking magnetic impurities ($k_B T_M=\hbar/\tau_M$ where $\tau_M$ is the transport relaxation rate), $v_F$ is the mean Fermi-velocity controlling the orbital limitation and $g$ the gyromagnetic ratio for the paramagnetic limitation. We defined: $T^{\star}(p) = \Omega\Psi(\lambda,\mu^{\star},T_M=0)$ as in reference \cite{Kos2003}. The functions $\Psi$ and $\Phi$ are calculated numerically as reported in \cite{Glemot1999}.

We assumed that pressure should take away the system from a magnetic QCP (with exception of the very low pressure range) decreasing both the pairing strength (reflected in a pressure dependence of $\lambda$) and the pair-breaking strength.
So we imposed the following constraints: we assumed that the change of \Hcslope{} for both field orientations should be entirely controlled by the pressure dependence of the strong coupling parameter, in other words that the pressure dependence of the Fermi velocities along the $\vec{c}$ and $\vec{a}$ axis follow $v_F^i = v_{F0}^i/(1+\lambda(p))$, $i=a,c$, with $v_{F0}^i$ constant. This implies a rather large initial value of the strong coupling constant : \Hcslope{} strongly decreases in a narrow pressure range, hence $\lambda(p)$ should be large (at least at low pressure), in order to provide enough dynamics on $v_F$ to fulfill that constraint. We also expect that the pair breaking mechanism due to magnetic fluctuations should be suppressed when going away from the QCP. These fluctuations certainly also control the value of the residual resistivity. As the decrease of the residual resistivity with pressure saturates around $\sim 2$ GPa \cite{Nicklas2001,Sidorov2002}, we assumed that  $T_M$ is essentially zero at 4 GPa (most likely even earlier).  Eventually, we adjusted $T_M$ against the measured pressure dependence of  \Tc, assuming that $\Omega$ has no pressure dependence, and with $\lambda(p)$ controlled by the pressure dependence of \Hcslope. Pair breaking was not included in our previously published analysis \cite{Knebel2008} and all the other parameter let free to vary with pressure.

So only $\lambda$, $T_M$ and $g$ were allowed to vary with pressure, and only $g=g^i$, $v_{F0}^i$, $i=a,c$ were allowed to be different for both fields orientations. All the parameters are constant with magnetic field. A matter of debate could be whether the Fermi velocity is field dependent or not ? Indeed, at zero pressure, for fields above $H_{c2}(0)$, the Sommerfeld coefficient ($\gamma=C_P/T$) extrapolated to $T\rightarrow 0$ decreases with field, which indicates that the effective mass decreases, and hence the Fermi velocity increases \cite{Bianchi2003} ($v_F\propto 1/m^\star\propto 1/\gamma$). However, for fields below  \Hczero{}, there is no indication of a field dependence of the specific heat ($C_p$) in the normal state, for both fields orientations \cite{Ikeda2001, Bianchi2003FFLO}. 
There is an increase of $C_P/T$ at low temperatures, reflecting a low characteristic energy scale, but no field dependence at a given temperature, as if this low energy scale remains field independent. So the behaviour of the Fermi velocity with field appears to be asymmetric, flat below $\approx 5T$ for $H \parallel \vec{c}$ and increasing above that field. For $H \parallel \vec{a}$, the same field independence of the specific heat is observed below $H_{c2}(0)$ \cite{Bianchi2003FFLO}. For more discussions on the field dependence of the effective mass near a quantum critical point from numerous experimental examples see the recent review \cite{FlouquetJPCS2011}. Hence at ambient pressure, a variation of the effective mass is only expected close to the field induced QCP (4.8 Tesla) and the Fermi velocity should not vary in most of the field range of our fit ($\sim$0-4 Tesla), large enough to obtain the different parameters of our model. 
In addition, the orbital limitation, which is controlled by the Fermi velocity, only governs \Hc $(T)$ in the low fields range, whereas at the lowest temperatures, \Hc $(T)$ is controlled by the paramagnetic limitation. So even if there is an increase of the effective mass on approaching the field induced QCP (close to $H_{c2}(0)$), it will have only little influence on the upper critical field. 

Under pressure, there is no data allowing for an estimate of the field variation of the Fermi velocity, and in particular, we do not know if the change of the location of the QCP has an influence on this quantity.
So, rather than introducing additional uncontrolled parameters, we chose, like in all previous studies, not to consider any field dependence of the Fermi velocity.


We use an s-wave model in spite of the well established d-wave nature of the superconducting ground state of \CeCoIn{}. This has been done only for sake of simplicity. Indeed, all the effects we take into account: strong-coupling,  pair-breaking and paramagnetic effects, are well accounted for by such an s-wave model. It is known that the general shape of \Hc$(T)$ is little influenced by the order parameter symmetry: only fine details like a weak anisotropy (see for example \cite{Wang1998}), or the absolute value of \Hcslope, are sensitive to the microscopic state. As will be discussed at the end, the main interest of taking d-wave symmetry into account would be to distinguish between a mass renormalization parameter ($\lambda_\Sigma$) and a pairing parameter ($\lambda_\Delta$, see \cite {Monthoux2001}). However, for such a distinction to be relevant, we would also need to take account of a realistic Fermi surface and spin fluctuation mode. This is clearly beyond our scope and probably beyond the actual state of the art of the field.

The model calculates \Hc{} for a second order phase transition, whereas the transition is known to become first order at low temperatures, typically below $T_c/3$ \cite{Bianchi2002}, most likely due to the Pauli limitation. Let us note that in CeCoIn$_5$, the paramagnetic limitation is so strong that it is reflected in the temperature dependence of \Hc{} already close to \Tc : reproducing the strong curvature of \Hc{} observed between \Tc{} and $T_c/3$ already controls the value of the paramagnetic limitation. What may seem puzzling is that the fit remains accurate in the whole temperature range, even when the transition is first order. This is certainly related to the fact that our calculation includes self consistently the appearance of an FFLO phase \cite{Glemot1999}, so that the difference of \Hc{} between a first order transition and a second order transition in the FFLO phase is probably within the error bars of the experimental data. This is also supported by theoretical calculations, which find little differences between the calculated first order and second order lines for the FFLO state, when it could be calculated \cite{HouzetPhysicaC1999,HouzetPRB2001}.


\begin{figure}[hbt]
\begin{center}
\includegraphics[width=0.8\textwidth]{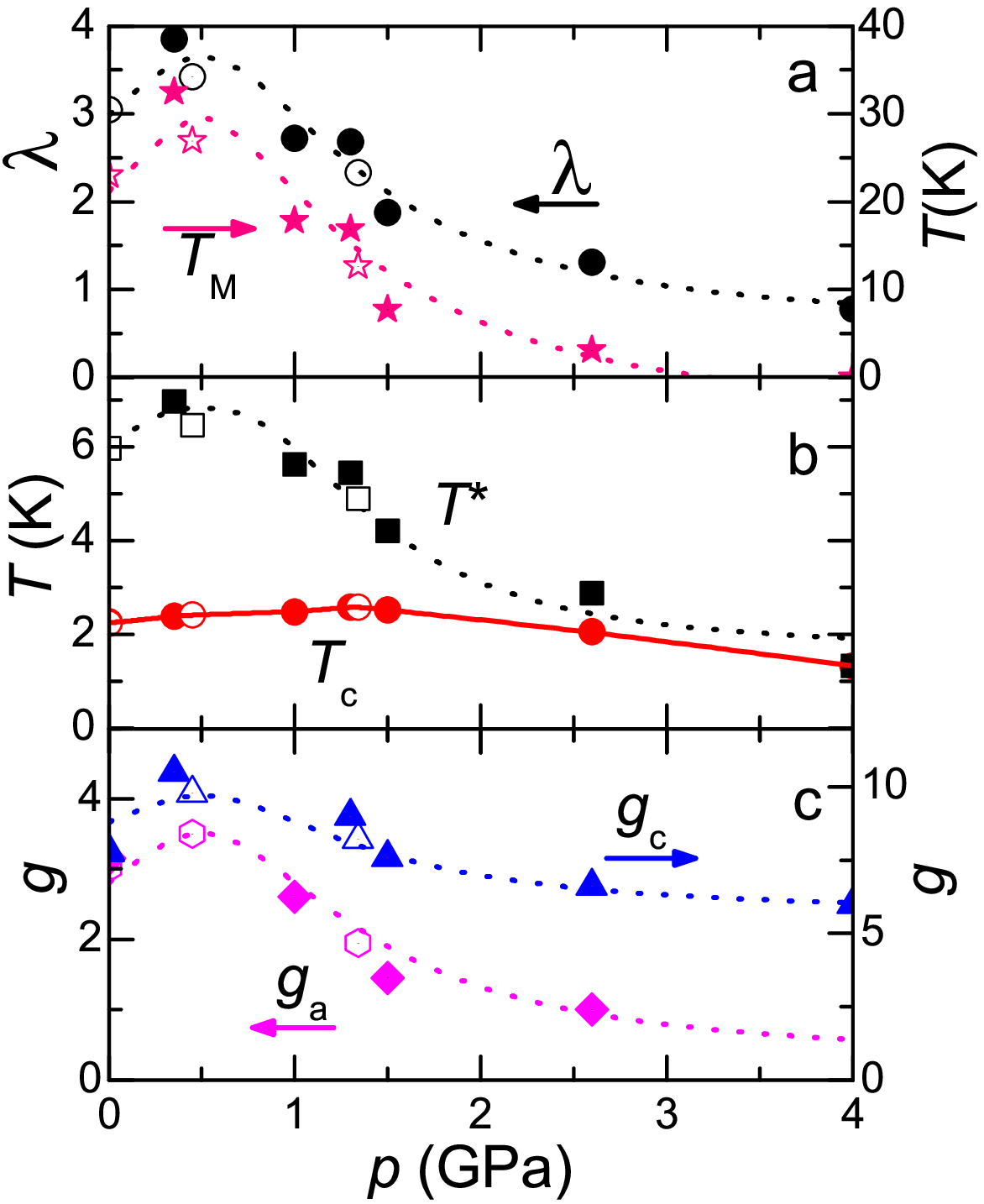}

\caption{\label{Parra}(colors online) a and c, parameters used for the fit of $H_{c2}(T,p)$:  all parameters have a maximum around 0.4 GPa, pointing to a QCP at this pressure, rather than at \pmax. (a) pressure evolution of $\lambda$ and of the pair breaking strength $T_M$. The variation from about 3.5 to 1 of $\lambda$ is controlled by the variation of $\partial H_{c2}(T_c)/\partial T$, for both field directions. (c) pressure evolution of the gyromagnetic ratio. (b) pressure evolution of $T_c$ (and $T^\star$), the superconducting transition at $H=0$ with (and respectively without) magnetic pair breaking, see text. Lines are guide for the eyes.}
\end{center}

\end{figure}

\section{Results and Discussion}
\begin{figure}[hbt]
\begin{center}
\includegraphics*[width=0.9\textwidth,bb=160 440 415 692]{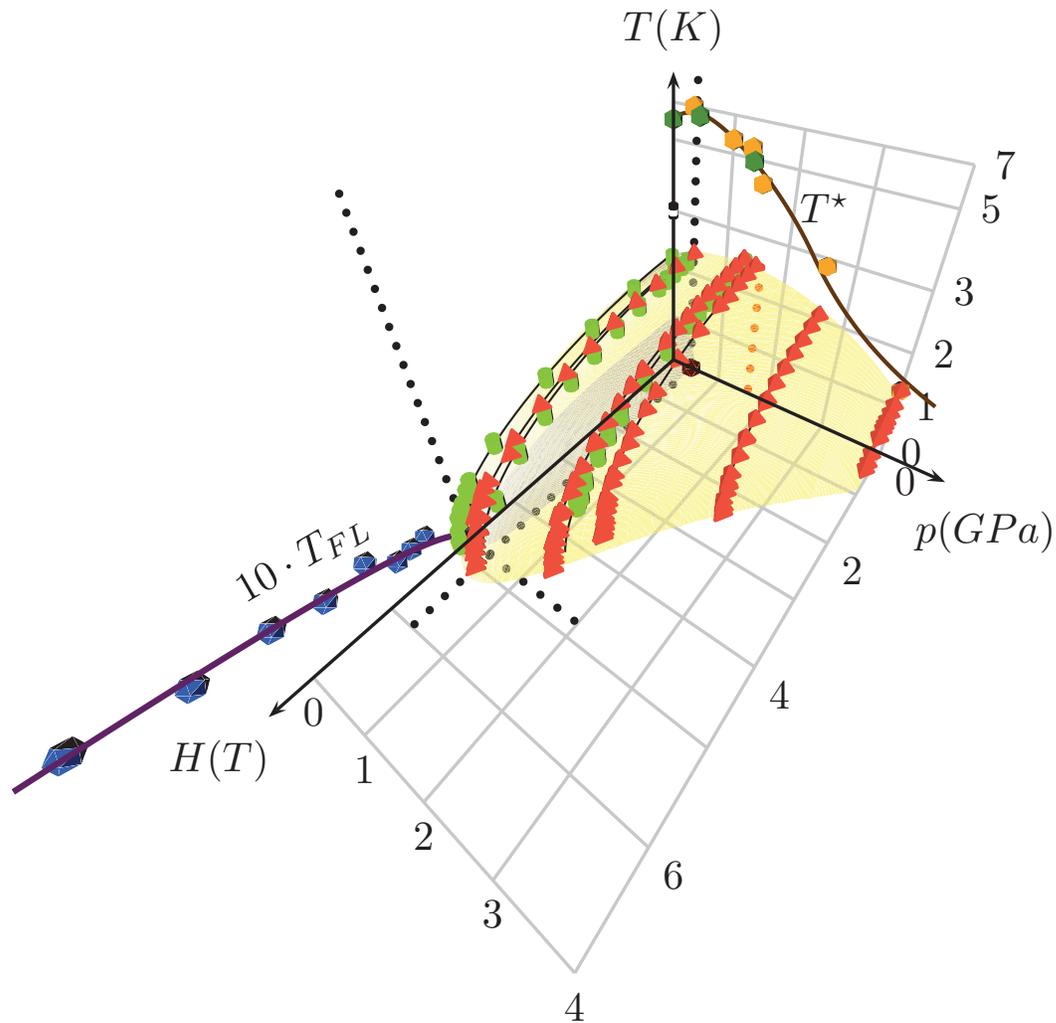}
\caption{\label{PhaseDiag} (colors online) New phase diagram proposed for \CeCoIn: the hypothetical QCP, corresponding to the maximum pairing strength as deduced from the analysis of \Hc{} (braun symbol on the pressure axis), is not at the maximum \Tc. Possible connection with the field induced QCP observed at zero pressure (end of the full line marked $10\cdot T_{FL}$ from \protect{\cite{Howald2011}}) is also displayed (gray surface, for a putative AFM ordered state in the absence of superconductivity). Green circles \protect{\cite{Miclea2006}} and red triangles \protect{\cite{Knebel2010}}: superconducting transition from specific heat. Green and orange hexagons, $T^\star$ line representing  the $T_c$ without pair-breaking in our model calculation (data of \protect{\cite{Miclea2006}} and \protect{\cite{Knebel2010}} respectively) : $T^\star$ is maximum at the pressure induced QCP. Dotted black lines give the pressures and fields values of the two measured QCP. Orange dotted line pressure of maximum of \Tc.
}
\end{center}
\end{figure}

We could find a set of parameters yielding quite satisfactory fits of \Hc : the fits are displayed, together with our data points, on figures \ref{GraHc}, as well as on figure \ref{GraHa}. On figure \ref{GraHa}, data of reference \cite{Miclea2006} are also displayed (data of \cite{Miclea2006} for $H\parallel c$ are only displayed on figure \ref{PhaseDiag}, for clarity, but equally well fitted).  In particular, we can see that the change of \Hcslope{} for both directions, can be well reproduced by the pressure dependence of a unique parameter $\lambda$ (also shown on figure \ref{Fig:slope}). The complete pressure dependence of the parameters used for the fit are displayed on figure \ref{Parra}. As expected $\lambda$ is essentially a decreasing function of pressure, except at very low pressure where it exhibits a maximum around $0.4$ GPa, imposed by the maximum of \Hcslope{} at this pressure.
Surprisingly, even this small detail is consistent with the NQR \cite{Yashima2004}, and the specific heat data \cite{Sparn2002}, which point respectively to a maximum of the gap to \Tc{} ratio and specific heat jump to \Tc{} ratio in the same pressure range. Therefore, this analysis of \Hc{} as well as the previous NQR and specific heat works, do suggest that the pairing strength is maximum neither at $0$ or negative pressure, nor at \pmax, but rather at $\approx 0.4$ GPa. 

This weak maximum of $\lambda$ suggests a QCP at  $\approx 0.4$ GPa, instead of the $1.3$ GPa corresponding to the maximum of \Tc. To our knowledge, only the work \cite{Ronning2006} attempted to locate the pressure induced QCP, with the pressure dependence of the field induced QCP in \CeCoIn{}, concluding that it would reach zero field around \pmax{}. Our analysis  is not in contradiction with the raw data of \cite{Ronning2006}, which show that the divergence of the $A$ coefficient of resistivity on approaching $H_{c2}$ is rapidly suppressed under pressure. Determination of the location of the putative QCP inside the superconducting dome with this technique necessarily implies large error bars. We should also note that the parameter controlling the magnetic pair breaking $T_M$ has a maximum at the same pressure value $\sim 0.4$ GPa, which supports the idea that the two mechanisms: interaction strength and magnetic pair breaking, originate from a unique effect associated with the QCP. The absolute value of $T_M$ might seem high compare to \Tc{}: it gives the mean free path of the quasi-particles $l=\hbar v_F/k_BT_M$ which has to be compared to the coherence length $l>\xi_0=\hbar v_F/\pi \Delta$. \Tc{}  is completely suppressed when $l\approx \xi_0$ (for magnetic impurities in the s-wave case or any kind of impurities for unconventional superconductors \cite{BalatskyRMP2006}), but with $\xi_0$ calculated without pair-breaking effect. This is why $T_M$ has to be compared to $T^\star$. Strong coupling effects also increase the ratio $\Delta/$\Tc, and so $\xi_0$ is even smaller than expected from weak-coupling formula when referred to \Tc. These effects are  fully included in our microscopic (strong coupling) model.


The last parameter of the model is the gyromagnetic factor. $g$ is most sensitive to the low temperature part of \Hc, where the transition becomes experimentally first order, and a field induced antiferromagnetic phase appear for $H\parallel \vec{a}$. Because our calculations are restricted to a second order phase transition (with FFLO phase), the value of g deduced from the fit could be less significant than that of $v_F$. Nevertheless, owing to the strong curvature of \Hc$(T)$ even close to \Tc, $g$ is already well determined within the limit of validity of the model, and the overall behavior with pressure is certainly correct. A striking feature, independent of the model, is the strong anisotropy of $g$, which points to a regime strongly different from the free electron case. Such an anisotropy, and the observed continuous decrease with pressure have also been observed on the bulk magnetic susceptibility \cite{Tayama2002}. Another one is the large value we deduce along the $\vec{c}$-axis, which results from the strong Pauli limitation in that direction, combined with the rather large value of $\lambda$ we need to fit the pressure variation of \Hcslope. However, theoretical predictions for magnetically mediated superconductivity show that one should distinguish, for non ``s-wave'' symmetry of the interaction, a strong coupling constant for the mass renormalization and for the pairing strength (respectively $\lambda_Z$ and $\lambda_\Delta$ in the notations of \cite{Monthoux2001}). Because the absolute value of $g$ is mainly governed by $\lambda_\Delta$, a more correct treatment of  \Hc{} for magnetically mediated superconductivity would lead to smaller values of $g$. $\lambda_\Delta<\lambda_Z$ and $\lambda_Z=\lambda$ is given by the variation of initial slope of \Hc. It would also imply less magnetic pair breaking ($T^\star$ depends on $\lambda_\Delta$).

Figure \ref{PhaseDiag} shows the phase diagram of \CeCoIn{} we can redraw on the basis of our measurements and analysis of \Hc{} for H$\parallel$c. We have obtained experimentally two points on the quantum critical line that would appear in absence of superconductivity. At p$=0.4$ GPa, H$=0$ in this work (for both field orientations although only one is shown here) and from our previous resistive measurements at p$=0$ GPa, H$=4.8$T (see \cite{Howald2011} and reference therein for this field induced QCP). 

\section{Conclusion}
Naturally, a proper quantitative analysis of the pressure dependence of \Hc{} requires a detailed knowledge of the pairing mechanism, and the associated theoretical modeling. But already at this semi-phenomenological level, we could unveil for the first time, the expected decoupling between optimum \Tc{} and maximum pairing strength. This decoupling is due to dominant pair-breaking effects in the neighborhood of the QCP, already pointed out previously by normal state features \cite{Nakatsuji2004,Nicklas2001} or anomalous superconducting properties \cite{Kos2003,Bang2004}. Probably due to a stronger coupling regime, or stronger 2D character \cite{Monthoux2001}, \CeCoIn{} is different from his parent CeRhIn$_5$, where the coincidence of the QCP and maximum \Tc{} is well documented. We claim that many peculiar features of \CeCoIn, like the large specific heat jump at \Tc{} and its pressure dependence, the pressure dependence of the gap to \Tc{} ratio observed by NQR, the pressure dependence of the paramagnetic limitation and of the initial slope of \Hc{} can be well explained in this scenario. This result provides an additional evidence on the link between superconductivity and quantum criticality. It strongly suggests that the pairing mechanism is the same as the one responsible for the appearance of a QCP. The magnetic origin of the QCP in \CeCoIn{} is likely, as demonstrated by magnetic fluctuations in the normal state and  proximity to an AFM phase. We hence believe that the pressure phase diagram of \CeCoIn{} stands as a paradigm of an (almost 2D), strongly coupled, antiferromagnetically mediated superconductor.

\section*{Acknowledgments}
We are pleased to thank J. Flouquet and Y. Yanase for stimulating discussions. This work was supported by the French ANR grants SINUS and DELICE. 

\section*{References}
\bibliographystyle{vancouver}


\end{document}